\documentclass[prl,twocolumn,superscriptaddress]{revtex4}

\usepackage{amsmath,amssymb,graphicx,color,bm,soul}

\usepackage{textcomp} %KWinkler added for \textmu support

\usepackage{hyperref}

\hypersetup{
    colorlinks=true,       % false: boxed links; true: colored links
    linkcolor=blue,          % color of internal links
    citecolor=blue,        % color of links to bibliography, !!!für pdf darkblue!!!
    urlcolor=black,           % color of external links
    breaklinks=true                                                          % Links können Zeilenumbruch aufweisen
}

\begin{document}

\title{Collective state transitions of exciton-polaritons loaded into a periodic potential}
\author{K. Winkler}
\affiliation{Technische Physik, Wilhelm-Conrad-R\"ontgen-Research Center for Complex
Material Systems, Universit\"at W\"urzburg, Am Hubland, D-97074 W\"urzburg,
Germany}

\author{O. A. Egorov}
\affiliation{Institute of Condensed Matter Theory and Solid State Optics, Abbe
Center of Photonics, Friedrich-Schiller-Universit\"at Jena, Max-Wien-Platz 1, 07743 Jena, Germany}

\author{I. G. Savenko}
\affiliation{National Research University of Information Technologies, Mechanics and Optics (ITMO University), Saint-Petersburg 197101, Russia}
\affiliation{COMP Centre of Excellence at the Department of Applied Physics, P.O. Box 11000, FI-00076 Aalto, Finland}

\author{X. Ma}
\affiliation{Institute of Condensed Matter Theory and Solid State Optics, Abbe
Center of Photonics, Friedrich-Schiller-Universit\"at Jena, Max-Wien-Platz 1, 07743 Jena, Germany}

\author{E. Estrecho}
\affiliation{Nonlinear Physics Centre, Research School of Physics and Engineering, The Australian National University, Canberra ACT 2601, Australia}

\author{T. Gao}
\affiliation{Nonlinear Physics Centre, Research School of Physics and Engineering, The Australian National University, Canberra ACT 2601, Australia}

\author{S. M\"uller}
\affiliation{Technische Physik, Wilhelm-Conrad-R\"ontgen-Research Center for Complex
Material Systems, Universit\"at W\"urzburg, Am Hubland, D-97074 W\"urzburg,
Germany}

\author{M. Kamp}
\affiliation{Technische Physik, Wilhelm-Conrad-R\"ontgen-Research Center for Complex
Material Systems, Universit\"at W\"urzburg, Am Hubland, D-97074 W\"urzburg,
Germany}

\author{T. C. H. Liew}
\affiliation{Division of Physics and Applied Physics, Nanyang Technological University
637371, Singapore}

\author{E. A. Ostrovskaya}
\affiliation{Nonlinear Physics Centre, Research School of Physics and Engineering, The Australian National University, Canberra ACT 2601, Australia}

\author{S. H\"ofling}
\affiliation{Technische Physik, Wilhelm-Conrad-R\"ontgen-Research Center for Complex
Material Systems, Universit\"at W\"urzburg, Am Hubland, D-97074 W\"urzburg,
Germany}
\affiliation{SUPA, School of Physics and Astronomy, University of St Andrews, St Andrews
KY16 9SS, United Kingdom}

\author{C. Schneider}
\affiliation{Technische Physik, Wilhelm-Conrad-R\"ontgen-Research Center for Complex
Material Systems, Universit\"at W\"urzburg, Am Hubland, D-97074 W\"urzburg,
Germany}

\begin{abstract}
{We study the loading of a nonequilibrium, dissipative system of composite bosons -- exciton polaritons -- into a one dimensional periodic lattice potential. Utilizing momentum resolved photoluminescence spectroscopy, we observe a transition between an incoherent Bose gas and a polariton condensate, which undergoes further transitions between different energy states in the band-gap spectrum of the periodic potential with increasing pumping power. We demonstrate controlled loading into distinct energy bands by modifying the size and shape of the excitation beam. The observed effects are comprehensively described in the framework of a nonequilibrium model of polariton condensation. In particular, we implement a stochastic treatment of quantum and thermal fluctuations in the system and confirm that polariton-phonon scattering is a key energy relaxation mechanism enabling transitions from the highly nonequilibrium polariton condensate in the gap to the ground band condensation for large pump powers.}
\end{abstract}

\maketitle

{\em Introduction.---} Since the initial demonstration of Bose-Einstein condensation of microcavity exciton polaritons~\cite{Kasrpzak2006}, one particular direction of research has been focused on engineering the potential landscape of polaritons. These potentials have been created by deposition of metal on the microcavity surface~\cite{Lai2007}, interference of surface acoustic waves~\cite{CerdaMendez2010}, deep etching of micropillars~\cite{Jacqmin2014}, fabrication of shallow mesas~\cite{Winkler2015}, and deposition of semiconductor micro-rods on a silicon grating~\cite{Zhang2015}. Periodic arrangements of these potentials have been used to study the formation of gap solitons~\cite{Tanese2013,CerdaMendez2013} and graphene-like band spectra (including Dirac cones)~\cite{Kim2013,Jacqmin2014}, are predicted to be prime candidates for the generation of topological polariton states~\cite{Karzig2014,Bardyn2015,Nalitov2015}, and hold high promise for implementation of quantum simulators in solid state systems~\cite{Bloch2012}.

Controlled loading of polaritons into a particular energy state of a band-gap structure enforced by the periodic potential remains a critical problem in this field. A polariton condensate is a nonequilibrium many-body system and therefore may form in an excited state~\cite{Maragkou2010}. Furthermore, polariton-polariton scattering, interactions with higher energy excitons, as well as scattering with phonons \cite{Wouters_relaxation,Savenko_relaxation}, can change the occupation of different energy states and lead to effective energy relaxation. For example, previous investigations of shallow 1D periodic potentials have demonstrated condensation initially occurring at the edges (high in-plane momentum) of the Brillouin zone (BZ) but then relaxing to the zero in-plane momentum state of the ground Bloch band with increasing pump power~\cite{Lai2007, Kim2011}. In general, experiments with polariton condensates in periodic potentials under non-resonant optical excitation rely on serendipity rather than on a controlled process to obtain a condensate in a particular energy and momentum state~\cite{Kim2011,Kim2013,Jacqmin2014}. In order to progress towards desired control over the loading process, it is important to understand the mechanisms for the condensate formation and transition between energy states.

In this Letter, we develop a profound understanding of these nonequilibrium effects, by carrying out a combined experimental and theoretical investigation of the polariton condensation in a one-dimensional array of buried mesa traps~\cite{Daif2006,Winkler2015} that form an effective {\em tight-trapping} periodic potential with the depth of several meV. We demonstrate that condensation of exciton-polaritons in a periodic tight-trapping potential exhibits two qualitatively different regimes: First, in the regime of moderate pump powers above the threshold of polariton condensation, exciton-polaritons condense into excited states with the energy blueshifted into the gap of the linear band-gap spectrum. Real- and momentum-space structure of these gap states is uniquely determined by the geometry of excitation. In particular, we observe controllable excitation of stable on- and off-site {\em discrete gap states} \cite{Kivshar2003,Louis2003} by on-site and off-site illumination with a tightly focused pump spot, respectively. This is in contrast to previous studies of more shallow lattices, where, under similar excitation conditions, the outcome was determined by markedly different stability properties of the gap states~\cite{Tanese2013}. Secondly, for pump powers {\em an order of magnitude} above the condensation threshold, the condensate relaxes to the lowest energy state with a significant redshift towards the ground-state band. The observed behavior cannot be captured by theoretical description in the limit of thermal equilibrium~\cite{Byrnes2010}. We therefore extend a nonequilibrium model~\cite{RefWoutersPRL991404022007,Ostrovskaya2013} to include a stochastic treatment of fluctuations in the quantum driven-dissipative system~\cite{Wouters2008}, and achieve excellent qualitative and quantitative agreement between numerical modeling and experimental measurements for all accessible excitation conditions. In particular, we uncover the crucial role of phonon-induced polariton energy relaxation in this system, which is responsible for condensation into the lowest energy and momentum states.

\begin{figure}[htp]
  %\centering
  \includegraphics[width=0.47\textwidth]{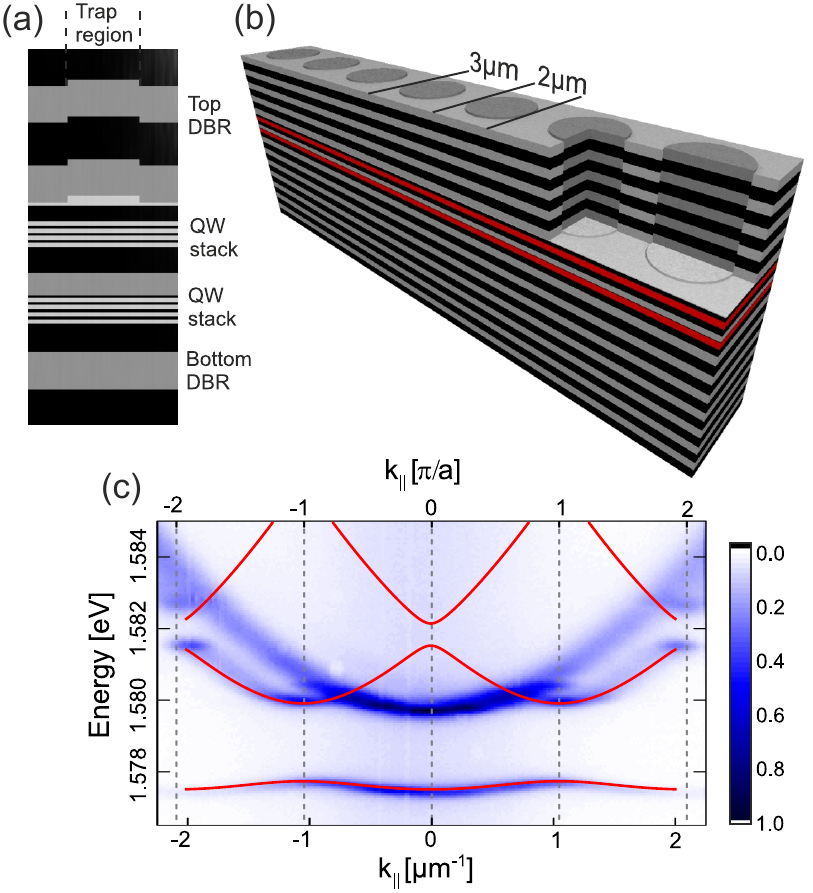}
  \caption{(color online) Schematics of the semiconductor microcavity: (a,b) Details of the textured cavity region (see text); (c) Angle-resolved photoluminescence spectra below condensation threshold shows formation of a band-gap polariton spectrum resulting from evanescent coupling between mesa traps. Red lines show calculated single-particle energy bands in the effective periodic potential of the depth $4.5$ meV. The measurements are recorded at a detuning between the cavity photon and exciton energies of $\Delta=E_C-E_X=-14$ meV (at $k=0$ of the ground Bloch band), which is comparable to the Rabi splitting, $E_R=11.5$ meV.}\label{Fig:1}
\end{figure}

\textit{Experiment.---}
We create a periodic lattice potential for polaritons by fabricating a one-dimensional array of attractive polariton potentials, coupled via the evanescent optical field. The individual polariton traps are created by a well controlled local elongation of the cavity layer thickness~\cite{Daif2006}. The MBE-grown layered structure [see Fig.~\ref{Fig:1}(a) and (b)] of the microcavity consists of an AlAs-$\lambda/2$-cavity surrounded by 37 (32) bottom (top) AlGaAs/AlAs DBR mirror pairs while the active media comprises two stacks of $7$ nm GaAs quantum wells distributed at the optical antinodes inside the cavity layer and at the first DBR interface. By patterning the $10$ nm thick GaAs-layer on top of the cavity layer in an etch-and-overgrowth step, we introduce an attractive photonic potential with a height-difference of about $5$ nm which amounts to about $5$ meV. More details on the process can be found in~\cite{Winkler2015}. A one-dimensional periodic array of 40 circular mesas with 2 \textmu m diameter and the period of $a = 3$ \textmu m is depicted in Fig.~\ref{Fig:1}(b). As exciton-polaritons are hybrid light-matter quasiparticles, the photonic potential yields a trapping potential for polaritons~\cite{Boiko2008,Kaitouni2006}, which also supports polariton condensation~\cite{Winkler2015}.

To study polariton condensation in the one-dimensional array, we perform angle-resolved photoluminescence measurements under continuous wave non-resonant excitation of the polariton states with a focused spot ($\sim 3.8$ \textmu m FWHM) of a frequency tunable Ti:Sa-laser. Fig.~\ref{Fig:1}(c) shows the far field emission spectra of such an array. The observed emission {\em below the condensation threshold} strongly differs from that of an isolated trap, where the trapped modes are discrete in energy and momentum space~\cite{Kaitouni2006}. Evanescent coupling of the modes confined in the polariton traps leads to formation of a distinct band-gap spectrum due to the spatial periodicity of the structure. Up to three Bloch bands are visible in the spectrum of the lower polariton, as shown in Fig.~\ref{Fig:1}(c). Because of the deep confinement, a gap between the ground and the first excited Bloch band evolves as large as $2.3$ meV. The experimentally determined band-gap structure of the spectrum can be fitted by a band-gap spectrum of single-particle eigenstates in a periodic one-dimensional lattice~\cite{Winkler2015} [see Fig.~\ref{Fig:1}(c)]. In addition to the bands resulting from the periodic potential for polaritons, the characteristic parabolic dispersion that stems from the planar, mesa-free region around the traps is also visible in Fig.~\ref{Fig:1}(c).

\begin{figure}[htp]
  %\centering
  \includegraphics[width=0.47\textwidth]{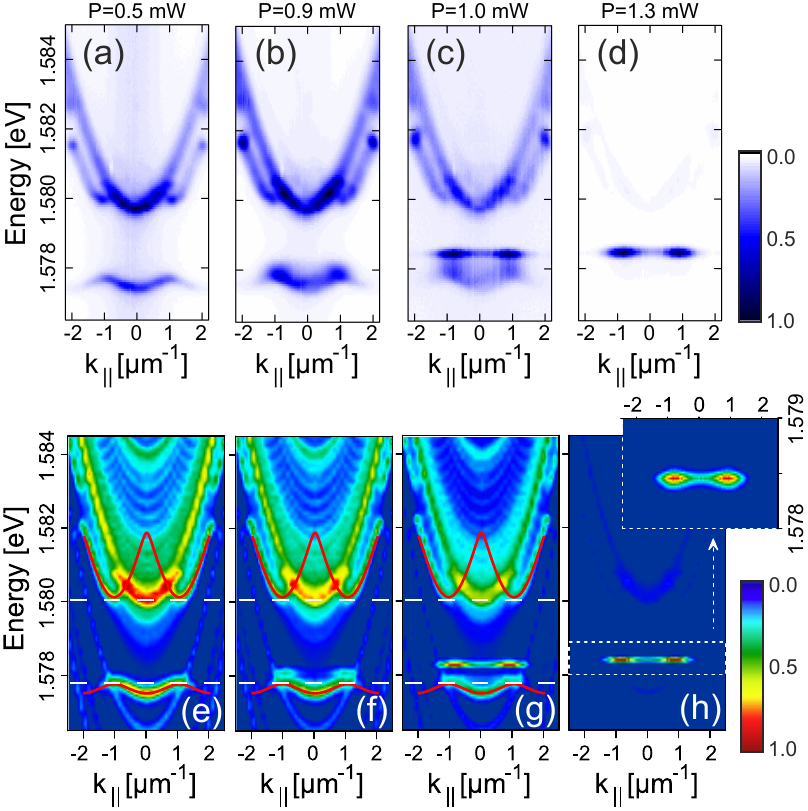}\\
  \caption{(color online) Experimental (a-d) and theoretical (e-h) study of the condensation in the one-dimensional array of mesas for an excitation with a narrow pump beam focused on a mesa ("on-site" excitation regime). (a-d) Far-field photoluminescence spectra recorded at various pump powers. (e-h) The dependence of polariton occupation at different energies as a function of in-plane momenta, $k_{\parallel}$, calculated using~(\ref{EqGPEnR}) and~(\ref{EqGPEnR2}). Large/small occupation numbers correspond to strong/weak intensity of the cavity emission in (a-d). Calculated occupation is integrated over the transverse-to-array coordinate, $y$. Shown are: (e) and (f) emission below the condensation threshold at the pump rates $P_0=1$ $\mu$m$^{-2}$ps$^{-1}$ and $P_0=10$ $\mu$m$^{-2}$ps$^{-1}$, respectively; (g) formation of the polariton condensate at the boundaries of the BZ in the vicinity of the threshold at $P_0=16$ $\mu$m$^{-2}$ps$^{-1}$; (h) condensation above the threshold at $P_0=20$ $\mu$m$^{-2}$ps$^{-1}$. The pump has a Gaussian profile with the FWHM of (a-d) $3.8$ $\mu$m and (e-h) $5$ $\mu$m accounting for exciton diffusion.}
 \label{Fig:2}
\end{figure}

Figures~\ref{Fig:2}(a-d) depict the evolution of the energy-momentum dispersion relation as a function of the pump power for {\em on-site} (i.e., focused on a {\em potential minimum}) excitation of the array. As the power of the laser excitation grows, the two first Bloch bands, as well as the planar cavity lower polariton mode, become more populated, as seen in Fig.~\ref{Fig:2}(b). Once the condensation threshold is reached at $\sim 1$ mW (see Supplemental Material (SM)~\cite{SupplementalMaterial} for details), and in agreement with previous experiments carried out in significantly more shallow potentials~\cite{Lai2007}, a polariton condensate forms at the vicinity of the BZ edges $k_{||}=\pm \pi/a$ deeply in the gap of the structure, with an approximate energy offset of $\approx 0.5$ meV with respect to the maximum of the ground band [Fig.~\ref{Fig:2}(c)]. With increasing power of the pump laser, the energy of the condensate shifts further into the gap, while remaining localized at the edges of the BZ in the momentum space.

The observed behavior of the condensed state is typical of a spatially localized polariton gap state reported in~\cite{Tanese2013}, and is reminiscent of a gap soliton well studied in the context of photonic waveguide arrays \cite{Kivshar2003} and atomic Bose-Einstein condensates in optical lattices~\cite{Oberthaler2004,Louis2003,Matuszewski2006}. Similarly to the matter-wave case, a polariton gap state is a defect mode, which shifts into the gap due to the repulsive nature of the inter-particle interactions~\cite{Yang2007}. The important aspect of the polariton in-gap condensation, pointed out in a previous study~\cite{Ostrovskaya2013}, is that this blueshift is caused by both the polariton-polariton interaction and interaction with the exciton reservoir.

Importantly, an {\em off-site} pumping results in a significantly different phenomenological behavior at any moderate pump power. Namely, we observe formation of an {\em off-site} gap state with two well pronounced peaks in real-space and centered around $k_{\|}=0$ in momentum space ( SM~\cite{SupplementalMaterial}). Therefore, in contrast to~\cite{Tanese2013}, the structure of the gap state in our tight-trapping lattice is determined solely by the excitation geometry.

\begin{figure}[t!]
\includegraphics[width=0.47\textwidth]{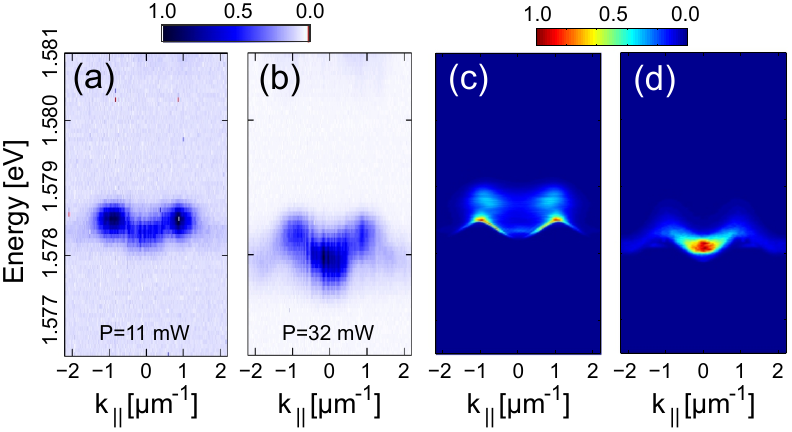}
\caption{(color online). Experimental and numerical spectra of polariton condensate in the high-power limit. (a,b) measured exciton-polariton dispersions for pump powers of $11$ and $32$ mW, respectively; (c,d) dependence of particle occupation at different energies as a function of in-plane momentum, $k_{\parallel}$, calculated using~(\ref{EqGPEnR}) and~(\ref{EqGPEnR2}) at the pump rates (c) $P_0=30$ and (d) $P_0=40$ $\mu$m$^{-2}$ps$^{-1}$. }
\label{Fig:3}
\end{figure}

At a higher power of about $10$ times the threshold [see Fig.~\ref{Fig:3}(a)], the condensation at $k_{||}=0$~$\mu m^{-1}$ becomes more prominent, and thermal effects begin to lead to a redshift of the emission towards the ground state of the system. At the highest excitation densities [$\sim 30$ times above threshold, Fig.~\ref{Fig:3}(b)], we observe a pronounced growth in the occupation of the energetic ground state of the system, which was previously reported in experiments carried out in very shallow trapping potentials~\cite{Lai2007}. So far, a theoretical understanding of the mechanisms responsible for the observed transition towards the ground state condensation has been lacking.

As well as a substantial redshift, the above transition is accompanied by increase in the linewidth and a growing population of lattice-free polaritons. Both of these issues can be overcome by  a {\em targeted} loading into the lowest-energy Bloch state of the system. As suggested in \cite{Ostrovskaya2013}, this can be achieved by an elliptically shaped pump beam with a large aspect ratio matched to the aspect ratio of the mesa array. The experimental results demonstrating controlled formation of the polariton condensate in the ground Bloch state at $k_{||}=0$ are shown in the SM ~\cite{SupplementalMaterial}.

\textit{Theory.---} Our experimental findings can be accurately modelled using a nonequilibrium Gross-Pitaevskii approach~\cite{RefWoutersPRL991404022007} with additional stochastic terms accounting for fluctuations in the exciton-polariton system~\cite{Wouters2008} and acoustic phonon-assisted relaxations~\cite{RefSavenkoPRL2013}. Such a stochastic approach can be rigorously derived within the truncated Wigner approximation~\cite{Wouters2009} and allows for the adequate description of the spontaneous coherence formation in the presence of the quantum and thermal noise. This approach provides a convenient theoretical tool for characterization of the photoluminescence from the semiconductor microcavity and, in particular, enables a theoretical description of the effects in the vicinity of the condensation threshold beyond the basic mean-field theory~\cite{RefWoutersPRL991404022007,Ostrovskaya2013}.

The stochastic mean-field equations for the polariton macroscopic wavefunction, $\psi(\mathbf{r})$, where $\mathbf{r}$ is a 2D coordinate vector, and the reservoir occupation number, $n_\textrm{R}(\mathbf{r})$, read~\cite{Wouters2008, Wouters2009, RefSavenkoPRL2013}
\begin{eqnarray}
\label{EqGPEnR}
i\hbar\frac{\partial\psi(\mathbf{r})}{\partial t}&=&{\cal F}^{-1}\left[E_\mathbf{k}\psi_\mathbf{k}+{\cal S}_\mathbf{k}(t)\right] +i\Gamma (n_\textrm{R})\psi(\mathbf{r})\\
\nonumber
&&+\left[V(\mathbf{r})+g_c|\psi({\mathbf{r}})|^2\right] \psi(\mathbf{r}) \\
\nonumber
&&+ i\hbar\frac{d\psi_\textrm{st}(\mathbf{r})}{d t} +\sum_\mathbf{k}\left[{\cal T}_{-\mathbf{k}}(t)+{\cal T}^*_\mathbf{k}(t)\right]e^{-i\mathbf{k}\mathbf{r}}; \\
\label{EqGPEnR2}
\frac{\partial n_\textrm{R}(\mathbf{r})}{\partial t}&=&-(\gamma_\textrm{R}+R|\psi(\mathbf{r})|^2)n_\textrm{R}(\mathbf{r})+P_0(\mathbf{r}),
\end{eqnarray}
where $\psi_\mathbf{k}$ is the Fourier transform of $\psi(\mathbf{r})$, ${\cal F}^{-1}$ stands for the inverse Fourier transform, $E_\mathbf{k}$ is the energy of free polariton dispersion. This kinetic energy is characterized by the effective mass of the lower polariton, $m$, which was taken as $4.5\times10^{-5}$ of the free electron mass. The periodic array of mesa traps is accounted for by the {\em two-dimensional} potential $V(\mathbf{r})$ .

The terms ${\cal S}_\mathbf{k}(t)$ and ${\cal T}_{-\mathbf{k}}(t)$ account for the acoustic phonon-mediated relaxation phenomena and are fully described in~\cite{RefSavenkoPRL2013}. We assume that phonons represent an incoherent thermal bath and use the Markov approximation~\cite{Carmichael} for their description. The term ${\cal S}_\mathbf{k}(t)$ describes the emission of phonons by the polariton system stimulated by the polaritons. The stochastic term ${\cal T}_\mathbf{q}(t)$ in the last line of Eq.~\eqref{EqGPEnR} describes spontaneous phonon-assisted decay. It is defined by the correlators~\cite{RefSavenkoPRL2013}
\begin{align}
\left<{\cal T}_\mathbf{q}^*(t){\cal T}_{\mathbf{q}^\prime}(t^\prime)\right>&=\left|G_{{\mathbf{q}}}\right|^2n_\mathbf{q}\delta_{\mathbf{q},\mathbf{q}^\prime}\delta(t-t^\prime);
\notag\\
\left<{\cal T}_\mathbf{q}(t){\cal T}_{\mathbf{q}^\prime}(t^\prime)\right>&=
\left<{\cal T}_\mathbf{q}^*(t){\cal T}_{\mathbf{q}^\prime}^*(t^\prime)\right>=0,
\label{EqThermal}
\end{align}
where $n_{\mathbf{q}}(T)$ is the temperature-dependent density of phonons in the state with the wavevector $\mathbf{q}$, $G_\mathbf{q}$ denotes the polariton--phonon scattering strength.

The reservoir (described by Eq.~\eqref{EqGPEnR2}), induces a net gain and energy blueshift through the real and imaginary parts of the term $\Gamma (n_\textrm{R})=\frac{\hbar}{2}(R n_R - \gamma_c)-i g_R n_R$ in Eq.~\eqref{EqGPEnR}, respectively. Here $R=0.001$ ps$^{-1}$~$\mu$m$^{2}$ defines the system--reservoir excitation exchange rate, $\gamma_c=0.33 \text{~ps}^{-1}$ and $\gamma_R=0.005$ ps$^{-1}$ are the decay rates of polaritons and reservoir excitons. The constants $g_c=10^{-3}$ meV$\mu$m$^{2}$ and $g_R=2 g_c$ characterize the strengths of polariton-polariton and polariton-reservoir interactions, respectively. The pump rate, $P_0(\mathbf{r})$, is induced by an off-resonant Gaussian-shaped laser beam with a $5$ $\mu$m FWHM.
Here the slightly broader pumping spot (the experimental value is 3.8 $\mu$m) accounts for diffusion of carriers in the conductivity band. Indeed simple estimations show that the electron-hole plasma created by a optical beam of 3.8 $\mu$m broadens towards 5 $\mu$m for a typical set of semiconductor parameters (i.e. for carriers life-time about 1 ns and for the diffusion about $D\simeq5$~cm$^{2}$~s$^{-1}$~\cite{RefWoutersPRL991404022007}).

The term
\begin{eqnarray}
\label{EqStochTerm}
d \psi_\textrm{st}(\textbf{r}_i)=\sqrt{ \frac{\gamma_c+R n_\textrm{R}(\textbf{r}_i)}{4\delta x \delta y} } dW_i,
\end{eqnarray}
in Eq.~\eqref{EqGPEnR} accounts for fluctuations by introducing a white noise into the model. Both the damping and gain, $\gamma_c$ and $R n_\textrm{R}$, contribute to this term~\cite{Wouters2008,Wouters2009}. Here $dW_i$ is a Gaussian random variable characterised by the correlation functions $\langle dW^{*}_idW_j\rangle=2\delta_{i,j}dt$ and $\langle dW_idW_j\rangle=0$
where $i$, $j$ are indices counting discrete mesh points $\mathbf{r}_i$ with the discretizations $\delta x$ and $\delta y$ in $x$ and $y$ directions, respectively.

\textit{Results and discussion.---}
First, we model the dynamics of the system at moderate pumping powers near the condensation threshold.
In this regime, the stimulated term ${\cal F}^{-1}{\cal S}_\mathbf{k}(t)$ is negligible compared to the fluctuations $i\hbar d \psi_\textrm{st}(\textbf{r}_i)/dt$, hence we can neglect the interaction with acoustic phonons. In order to match the experimental spectra and study photoluminescence from different positions on the sample, the momenta-resolved spectra have been integrated over the $y$-direction which is transverse to the array, see Figs.~\ref{Fig:2}(e-h) [see also Figs.~S2 (a-d) of SM~\cite{SupplementalMaterial}].

The results of our numerical simulations for the {\em on-site} excitation of the array are presented in Fig.~\ref{Fig:2}(e-h) and demonstrate a very good agreement with the experiment. Indeed, below the condensation threshold, both the band-gap spectrum and the parabolic 'background' dispersion are visible [Fig.~\ref{Fig:2}(e)]. For a stronger pump, the emission with non-zero momenta around the BZ edges ($k_{||}=\pm \pi/a$) is clearly observed, in excellent agreement with the experiment [Fig.~\ref{Fig:2}(f-h)]. This spontaneous build up of the condensate is accompanied by the blueshift of the energy into the gap above the ground band of the array. Our theoretical analysis confirms that the blueshift is mostly caused by the reservoir--induced repulsive potential which is determined by the term $g_R n_R$. The gap state bifurcates from the upper edge of the band characterised by the "staggered" phase, and therefore inherits the characteristic $\pi$-phase shift between neighbouring density peaks \cite{Ostrovskaya2013},  as shown in the SM~\cite{SupplementalMaterial}. Furthermore, we find that formation of the condensate in both the on-site (near $k_{||}=\pi/a$ momentum) and off-site (near $k_{||}=0$ momentum) states is controlled purely by the on-site or off-site positioning of the pump spot, respectively. This behaviour is very robust in a wide interval of pump powers [see Fig.~\ref{Fig:2}(h) and Figs.~S2(b-d) of the SM~\cite{SupplementalMaterial}], which confirms the {\em enhanced stability} of the gap states in the tight-binding lattice, in contrast to the shallow confinement case \cite{Tanese2013}.

At pump powers well above the condensation threshold, phonon--mediated energy relaxation can no longer be neglected. Moreover, the stochastic term $i\hbar d \psi_\textrm{st}(\textbf{r}_i)/dt$ in this regime becomes negligible compared to the stimulated term ${\cal F}^{-1}{\cal S}_\mathbf{k}(t)$. Hence we can neglect the former, thus assuming that the energy relaxation is predominantly driven by the phonons. A consequence of this phonon relaxation mechanism, taken into account by Eqs.~\eqref{EqGPEnR} and~\eqref{EqGPEnR2}, is manifested in the behaviour shown in Fig.~\ref{Fig:3}(c,d), which is in excellent agreement with experimental observations [Fig.~\ref{Fig:3}(a,b)]. Here, we see a distinct transition from the formation of a polariton condensate at the vicinity of $k_{||}=\pm \pi/a$ towards condensation at $k_{||}=0 $. Indeed, at the highest pump power, phonon--assisted relaxation reproduces the macroscopic occupation of the ground state of the system [see Fig.~\ref{Fig:3}(d)] observed in the experiment. It should be noted , that in Figs.~\ref{Fig:3}(c,d) we considered a 1D structure, thus neglecting 2D spreading of particles in the ($y$-) direction transverse to the periodic potential. Although this simplification leads to some quantitative differences, for instance, in the pump rates (c.f. Fig.~\ref{Fig:3}), it provides a very good  qualitative description of the experimental phenomena.

\textit{Conclusions.---} We have performed a comprehensive experimental investigation of the exciton-polariton condensation in a tight-binding, quasi one-dimensional periodic potential. We have observed transitions of condensed states between different energy states, which are efficiently controlled via intensity and spatial alignment of the off-resonant optical pump. Furthermore, we have developed an advanced model taking into account quantum and thermal fluctuations in the system, as well as interactions with acoustic phonons. Our modelling of polaritons trapped in a periodic array demonstrates excellent agreement with the observed dynamics {\em both below and above condensation threshold}, for all pumping configurations. In particular, we have identified and explained two dramatically different regimes of polariton condensation at moderate and high pump powers, respectively. Namely, we have conclusively demonstrated that population of different energy states by the condensed polaritons is strongly determined by the energy relaxation due to fluctuations (at moderate powers) or phonons (at high powers).

Our work provides a complete conceptual understanding of the nonequilibrium condensation of exciton-polaritons in periodic potentials, and thus represents a significant progress towards controlled loading of the condensate into microstructured lattices for fundamental studies and possible applications.

\textit{Acknowledgements.---}
O.A.E. acknowledges financial support by the Deutsche Forschungsgemeinschaft (DFG project EG344/2-1) and by the EU project (FP7, PIRSES-GA-2013-612600) LIMACONA.
I.G.S. acknowledges financial support from the Academy of Finland through its Centre of Excellence Programs (Projects No. 250280 and No. 251748); the Government of Russian Federation, grant 074-U01; and the Dynasty Foundation. The calculations presented here were partly performed using supercomputer facilities within the Aalto University School of Science Science-IT project. E.E., T.G. and E.A.O. acknowledge support by the Australian Research Council, discussions with Michael Fraser, and assistance of Robert Dall with the experimental setup. The W\"urzburg group acknowledges financial support from the state of Bavaria. Assistance of Anne Schade, Jonas Ge{\ss}ler and Monika Emmerling during fabrication and patterning of the sample is gratefully acknowledged.

\end{document}